\documentclass[showpacs,preprintnumbers,amsmath,amssymb]{revtex4}
\usepackage{epsfig}
\usepackage{graphicx}% Include figure files
\usepackage{dcolumn}% Align table columns on decimal point
\usepackage{bm}% bold math

\def\ben{\begin{enumerate}}
\def\een{\end{enumerate}}
\def\bit{\begin{itemize}}
\def\eit{\end{itemize}}
\def\beq{\begin{equation}}
\def\eeq{\end{equation}}
\def\bea{\begin{eqnarray}}
\def\eea{\end{eqnarray}}
\def\bq{\begin{quote}}
\def\eq{\end{quote}}
\def \lsim{\mathrel{\vcenter
     {\hbox{$<$}\nointerlineskip\hbox{$\sim$}}}}

\def\gappeq{\mathrel{\rlap {\raise.5ex\hbox{$>$}}
{\lower.5ex\hbox{$\sim$}}}}
\def\lappeq{\mathrel{\rlap{\raise.5ex\hbox{$<$}}
{\lower.5ex\hbox{$\sim$}}}}

\def\Etm{E_T \!  \! \! \! \! \! \! /~~}

\def\tmg{\tau \to \mu \gamma}
\def\m3e{\mu \to e \bar{e} e}
\def\tmmm{\tau \to \mu \bar{\mu} \mu}
\def\temm{\tau \to e \bar{\mu} \mu}

\def\Htt{h \to \tau^+  \tau^- }
\def\Htm{h \to \tau^\pm  \mu^\mp }
\def\Hte{h \to \tau^\pm  e^\mp }

\def\a{\alpha}

\def\m{\mu}

\newcommand{\met}{\mbox{\ensuremath{\slash\kern-.7emE_{T}}}}

\begin{document}

\title{
LHC sensitivity to  $\bm{ h   \to \tau^\pm \mu^\mp}$  }% Force line breaks with \\

\author{ Sacha Davidson}
 \email{s.davidson@ipnl.in2p3.fr}
% \altaffiliation[Also at ]{Physics Department, XYZ University.}%Lines break automatically or can be forced with \\
\author{Patrice Verdier}%
 \email{verdier@ipnl.in2p3.fr}
\affiliation{%
 IPNL, Universit\'e de Lyon, Universit\'e Lyon 1, CNRS/IN2P3, 4 rue E. Fermi 69622 Villeurbanne cedex, France
}%

\begin{abstract}
We study the sensitivity of  the LHC, with 20 fb$^{-1}$  of data,  
to lepton flavour violating Higgs boson  decays $\Htm$ (and $\Hte$).
We consider the large population of Higgses  produced in
gluon fusion,  combined with leptonic decays of the $\tau$,
and estimate that the LHC could  set a  95\% confidence
level   bound  $BR(\Htm) < 
4.5 \times 10^{-3}$.
This correponds to a coupling of order the Cheng-Sher ansatz
$y_{\tau \mu} = \sqrt{m_\tau m_\mu/v^2}$.

\end{abstract}

% PACS, the Physics and Astronomy
                             % Classification Scheme.
%\keywords{Suggested keywords}%Use showkeys class option if keyword
                              %display desired
\maketitle

%%%%%%%%%%%%%%%%%%%%%%%%%%%%%%%%%%%%%%%%%%%%%%%%%%%%%%%%%%%%%%%%%%%%%%
%% INTRO        %%%%%%%%%%%%%%%%%%%%%%%%%%%%%%%%%%%%%%%%%%%%%%%%%%%%%%
%%%%%%%%%%%%%%%%%%%%%%%%%%%%%%%%%%%%%%%%%%%%%%%%%%%%%%%%%%%%%%%%%%%%%%

 A  new resonance, 
hereafter   refered to as the Higgs boson ($h$),   
was recently discovered \cite{LHCH} around 126 GeV.
Its couplings and branching ratios  will be determined
with additional data; meanwhile,  
one can  speculate on possible
departures from Standard Model (SM) expectations. It is known that
 Beyond-the-Standard-Model (BSM) physics is present 
in the lepton sector  to  generate neutrino masses, 
and  should induce Lepton Flavour Violating (LFV) Higgs decays
at some rate.  We suppose that this rate could be detectable.

    The prospects of observing 
LFV Higgs couplings have been studied in the past, both
at colliders \cite{LFVHcoll}, 
and in low energy experiments \cite{LFV} (see
also \cite{anna} for supersymmetric models, and
\cite{ACK} for a recent discussion with R-parity violation).
%{\bf Using Run~I data, the CDF experiment searched for a scalar particle
%decaying to $ \tau^\pm \mu^\mp$~\cite{CDF}.}
The low energy bounds, and their implications for
collider searches, were 
 studied in  detail for the Two Higgs Doublet
Model \cite{japonais,gerald}, and for a single
Higgs doublet  with effective operators
\cite{oleg2,HKZ,Chen:2006hp,MPR}
\footnote{For a clear review, see \cite{Ellisetal}.}.
They do not exclude observable $\Htm$ or $\Hte$ at the LHC.
{ The bounds that can be set from current $\Htt$ data,
and prospects of a dedicated $\Htm$ search 
(in the vector boson fusion channel) are
discussed in \cite{HKZ}.}

This brief note advocates
searching for $\Htm$  with Higgses  produced 
in gluon fusion, followed by  $\tau \to e \nu \bar{\nu}$.
This channel was discussed for heavy Higgs bosons
in \cite{DCGM}.
Due to the large  SM  gluon fusion  cross section,
which we denote as
$\sigma_{SM} (pp(gg) \to h) = 19.2 \pm 2.8$ pb,
this channel produces more signal events than
producing a Higgs in vector boson fusion (VBF),
$\sigma_{SM}( VBF)  =1.68 \pm 0.05$ pb.
The VBF channel is used for $h \to \tau^+ \tau^-$
because the two jets  distinguish the signal  in  
the large $(\gamma/Z)^* \to \tau^+ \tau^-$ background.
%(with  with 40 fb$^{-1}$ of data,
%$\sim$ 920 $h$s from VBF  decay to a hadronic tau and
%a leptonic tau). 
However, in the case of  $\Htm$,
 the hard $\mu$ momentum and specific
missing energy   allow the signal to be sufficiently
distinguished from the background that one obtains
an interesting sensitivity with current data.
Our sensitivity compares favourably with the
results of \cite{HKZ}, who studied $\Htm$
following Higgs production in VBF.

We are interested in the  lepton flavour violating Higgs decays 
$h \to \mu^\pm\tau^\mp$,
and $h \to e^\pm\tau^\mp$, because we expect
the ``SM-like'' Higgs to couple most strongly
to the third generation.  
We only study 
$h \to \mu^\pm \tau^\mp \to \mu^\pm e^\mp\nu\bar{\nu}$,
because we anticipate that the sensitivity to
$h \to e^\pm\tau^\mp \to e^\pm \mu^\mp\nu\bar{\nu}$ 
should be equivalent, or better.  The final
state particles 
 $e^\pm\mu^\mp + \Etm$  are the same,
so the  SM backgrounds should be similar, 
and the detection prospects for  a low $p_T$
muon from tau decay are better than those
of a  low $p_T$ electron.

The couplings of the SM Higgs are flavour diagonal,
so New Physics is required to allow $\Htm$.
For  a single  doublet  Higgs field $H$,
 the presence of non-renormalisable
operators, for instance  of the form 
$[H^\dagger H] [\overline{\ell}_\mu H] \tau_R$,
 allows
flavour-changing couplings for the physical Higgs
field $h$ \cite{BNGL}.  In the presence
of additional scalar fields, such as 
a 2 Higgs Doublet Model\cite{hhg}, or the
NMSSM \cite{ana, BEetal}, renormalisable  flavour-changing
interactions can  arise,
provided the mass spectrum is away from the decoupling limit
\footnote{For instance, in the 2HDM, two neutral scalars with
masses below the decoupling scale, is consistent
with low energy constraints including $b \to s \gamma$
\cite{gerald}.}.
  The production rate via gluon
fusion could  also be modified by
New Physics.  For instance, in the 2HDM,  neglecting
the heavy charged Higgses in the loop, it is
multiplied by $\sin^2 2(\beta-\alpha)$, where
$(\beta -\a )$ is the angle between the CP-even 
mass  eigenstate basis and the basis where only  one
doublet has a vev\footnote{A 2HDM with LFV is of ``Type III'',
so $\tan \beta$ is not uniquely defined \cite{DH}.
However, if the $\tau$ is the heaviest charged lepton,
and $\tan \beta_\tau$ in the ratio $y_\tau/m_\tau$,
where  
$y_\tau$  is the $\tau$ Yukawa coupling, then
the flavour-changing vertices have a factor
 $\tan \beta_\tau$. See \cite{gerald} for a
detailed discussion.}.

There are weak constraints on  LFV
Higgs decays from low energy data.
The limits from   processes 
which the Higgs could  mediate at tree level, 
such as $ \tmmm$, $ \temm$ and $\tau\to \eta \mu$,  allow ${\cal O}(1)$
LFV couplings to the Higgs \cite{japonais,paradisi,oleg2,HKZ}.
More restrictive model-dependent bounds arise from $\tmg$
(see {\it e.g.} \cite{oleg2,HKZ} for a single Higgs, and
\cite{japonais,paradisi,gerald} for the 2HDM); a LFV
Higgs coupling of order the tau yukawa is allowed.

We assume a single ``SM-like'' Higgs at 126 GeV.
In particular, we assume the change
in the total decay rate  due to New Physics 
is sufficiently small  that the narrow width approximation for
production and decay can be used.
Most of the backgrounds arise from non-Higgs
mediated processes,  so we study  the  observable
 \beq
\frac{\sigma(pp(gg)\to h)}{\sigma_{SM}(pp(gg)\to h)} BR(\Htm)
\eeq
This circumvents the issue
of parametrising New Physics couplings \footnote{see, {\it e.g.}
\cite{HxsecWG}, or \cite{aldo} for a recent discussion and
references.}.  However, in simulating
the backgrounds from Higgs decays, we use the SM
production rate $\sigma_{SM}(pp(gg) \to h)$. 
This is  innocuous because
the Higgs background is negligeable compared
to the dibosons and $Z^*$s.

%\section{Simulations+Analysis+Results}
%\label{simu}

%{\it
Our signal process is 
$pp(gg) \to h \to \tau^{\pm} \mu^{\mp} \to e^{\pm}\nu\bar{\nu} \mu^{\mp}$,
where the Higgs boson is produced by gluon fusion,
and was simulated with 
 the Monte Carlo event generator {\sc pythia} version~8.162~\cite{Sjostrand:2007gs}.
 Other {  production mechanisms  for the } Higgs boson were
not taken into account, since their rates are at least one order of magnitude lower. The Higgs 
boson mass was set to 126~GeV.

The background simulations from our previous study~\cite{DLV}
{ of lepton flavour violating $Z$ decays }
 was used
to estimate SM backgrounds leading to $e^{\pm}\mu^{\mp}$ dilepton final states.
These includes the following processes: $pp \to Z/\gamma^* \to \tau^+\tau^-$, 
$t\bar{t}$ and $tW$, and gauge boson ($W$ and $Z$) pair production.
In { the background simulations of} 
this new analysis, we also { included}  Higgs boson production
{ via} gluon fusion, { followed by lepton flavour
conserving decays}. The SM decays of the Higgs
boson leading to $e^{\pm}\mu^{\mp}$ final state are: $pp \to h \to \tau^{\pm}\tau^{\mp}$
where both tau lepton decay leptonically, $pp \to h \to W^+W^-$ where both $W$ 
decay leptonically, and $pp \to h \to ZZ^*$. Those processes were also simulated using
{\sc pythia} version~8.162 with a Higgs boson mass of 126~GeV.

The simulated signal and SM Higgs boson backgrounds are summarized in Table~\ref{tab1} { (see \cite{DLV} for similar information on other background processes)}.
The SM cross section of Higgs production by gluon fusion is 19.2~pb for a Higgs boson mass
of 126~GV with a total uncertainty of $\pm 14.7$~\%~\cite{Dittmaier:2011ti}. Branching ratios from SM were used
for $h \to \tau^{\pm}\tau^{\mp}$, $h \to W^+W^-$, and $h \to ZZ^*$ decays; those are
6.73\%, 23.3\%, and 2.85\%, respectively. 

Those Monte Carlo events were passed through the {\sc delphes} program~\cite{Ovyn:2009tx}, with the
anti-kt jet reconstruction of {\sc fastjet}~\cite{Cacciari:2011ma}, 
to simulate the CMS detector response as described in~\cite{DLV}.

\begin{table}[htb]
\begin{center}
\begin{tabular}{|c|l|r|r|}
\hline
\multicolumn{2}{|c|}{Processes}                          & $\sigma \times BR$ (pb) & Number of simulated \\
\multicolumn{2}{|c|}{\ }                                 &                                & events              \\
\hline\hline
SM Higgs back.   & $pp(gg) \to h \to \tau^{\pm}\tau^{\mp} \to \ell^{\pm}\nu\bar{\nu} \ell'^{\mp}\nu\bar{\nu}$ with $\ell,\ell'=e,\mu$ & 0.160    &  30,000 \\
SM Higgs back.   & $pp(gg) \to h \to W^+W^- \to \ell^+\nu \ell'^-\bar{\nu}$ with $\ell,\ell'=e,\mu,\tau$                              & 0.469    & 100,000 \\
SM Higgs back.   & $pp(gg) \to h \to ZZ^* \to f\bar{f}f'\bar{f}'$ with $f=q,\nu,e,~\mu,~\tau$                                         & 0.548    & 100,000 \\
\hline
Signal           &  $pp(gg) \to h \to \tau^{\pm} \mu^{\mp} \to e^{\pm}\nu\bar{\nu} \mu^{\mp}$                                             &          & 100,000 \\
\hline
\end{tabular}
\caption{\label{tab1}
Standard Model Higgs boson backgrounds in the search for 
lepton flavour  violating $h$ decay at the LHC  
(see \cite{DLV} for similar information about  non-Higgs processes);
the second column contains $\sigma \times BR$ (in pb) for p-p collisions at $\sqrt{s}=8~TeV$ for Higgs boson production via gluon fusion,
and the third column shows the number of simulated events in this analysis, 
which is in all cases 
greater than 10 times $\mathcal{L}\sigma BR$ with  $\mathcal{L}=20~fb^{-1}$.
}
\end{center}
\end{table}

The $h \to \tau^{\pm} \mu^{\mp}$ event selection follows closely what was done in~\cite{DLV}
to select $Z \to \tau^{\pm} \mu^{\mp}$ events with an electron and a muon back-to-back in the transverse
plane. Events are required to contain at least one muon with $p_T$ greater than
30~GeV and $|\eta|<2.1$, 
and at least one electron with $p_T$ greater than 15~GeV and $|\eta|<2.5$.
Events must contain exactly 2 opposite-sign (OS) leptons. 
Events having   one or more reconstructed jets
with $p_T>30$~GeV and $|\eta|<2.5$ are  rejected. 
The azimuthal angle between the muon and the electron,
$\Delta\phi(e,\mu)$, has to be higher than 2.7 radians. 
And the azimuthal angle between the electron and the 
direction of the missing transverse energy, $\Delta\phi(e,\met)$, 
has to be lower than 0.3.

\begin{figure}[htbp]
\begin{center}
\begin{tabular}{cc}
(a) & (b) \\
\includegraphics[width=8cm]{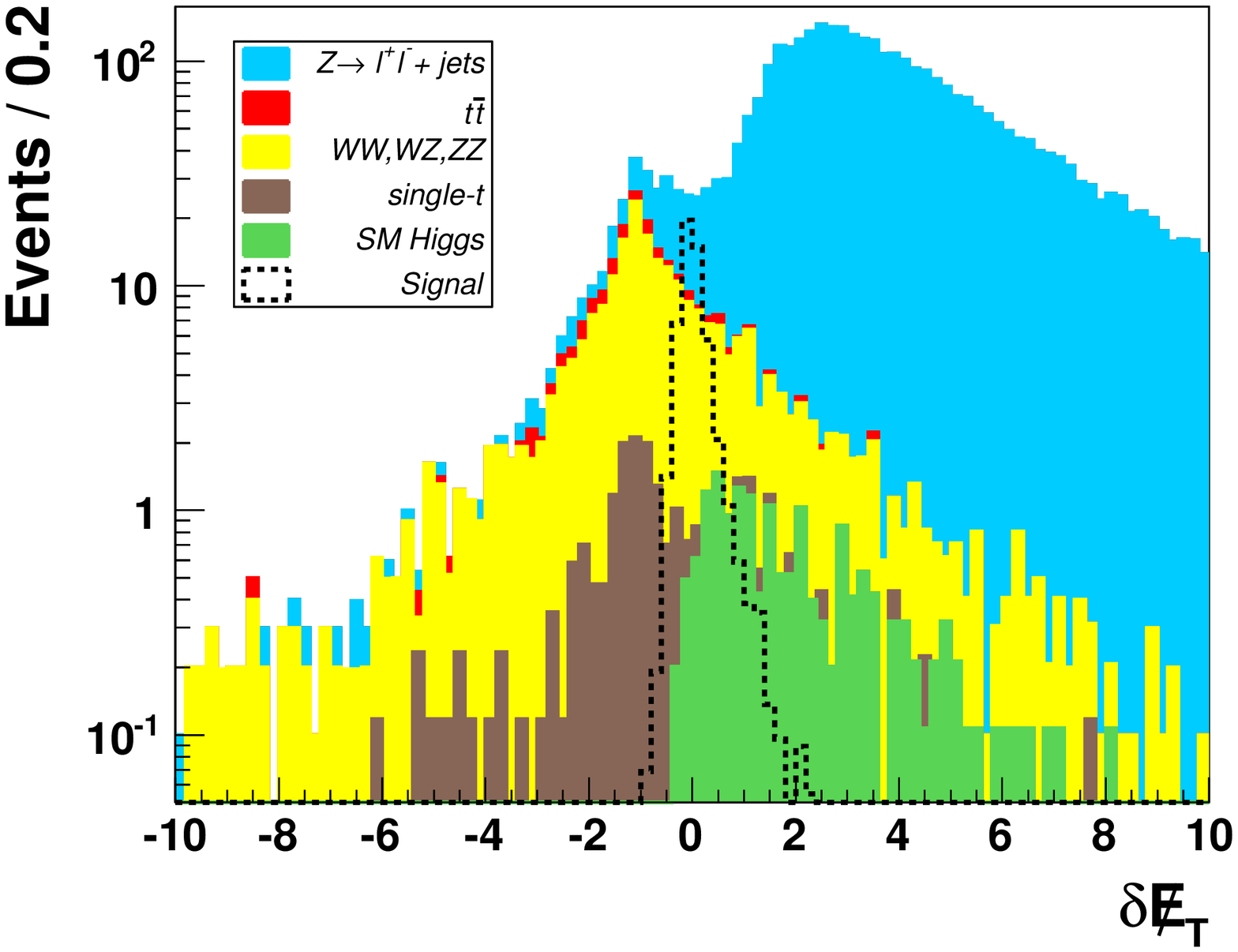} &
\includegraphics[width=8cm]{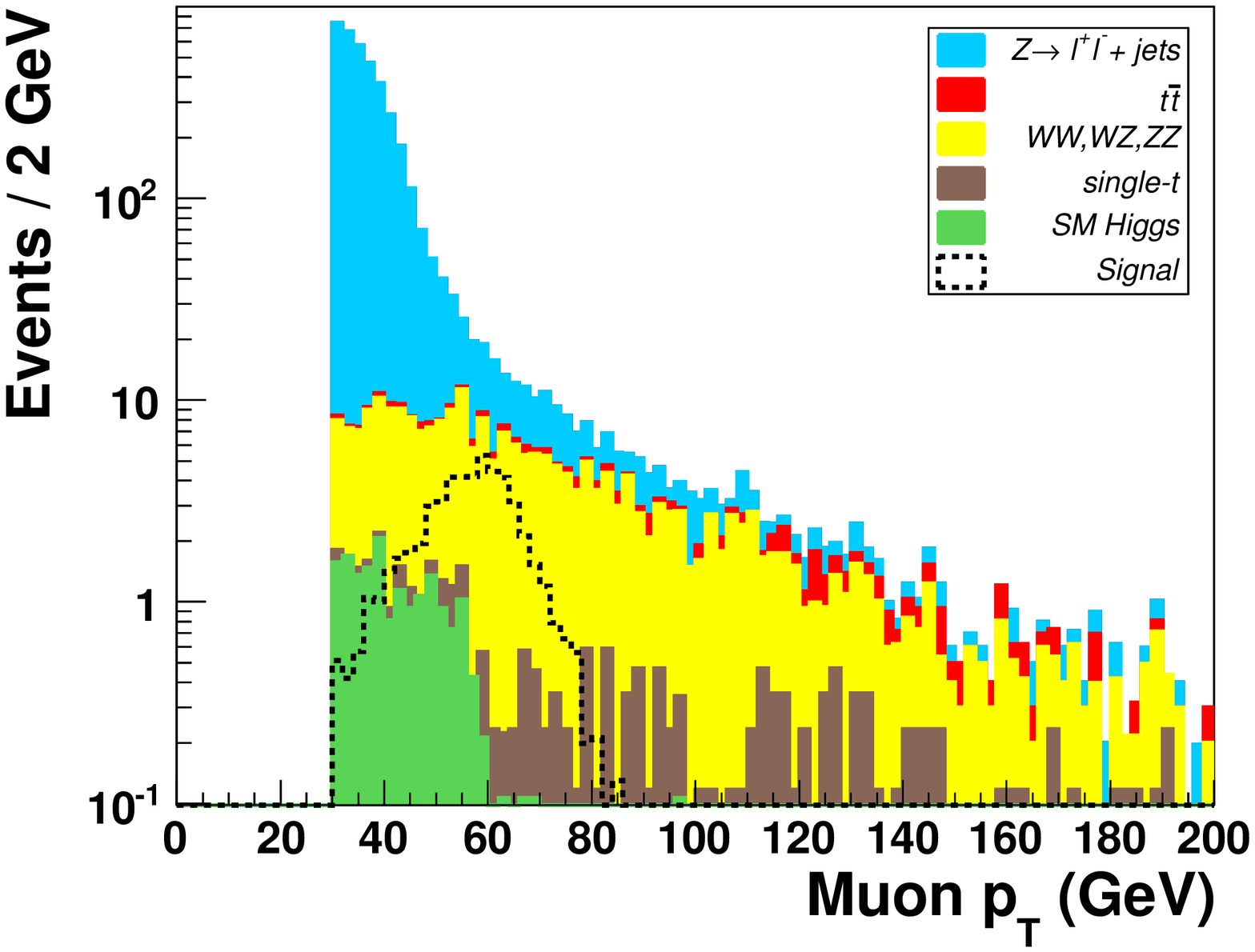} \\
\end{tabular}
\caption{
\label{fig1}
Distributions of $\delta\met$ (a), and of the muon $p_T$ (b), before applying the final selection criterium
in the $(\delta\met,p_T^\mu)$ plane.
}
\end{center}
\end{figure}

\begin{figure}[htbp]
\begin{center}
\begin{tabular}{cc}
(a) & (b) \\
\includegraphics[width=8cm]{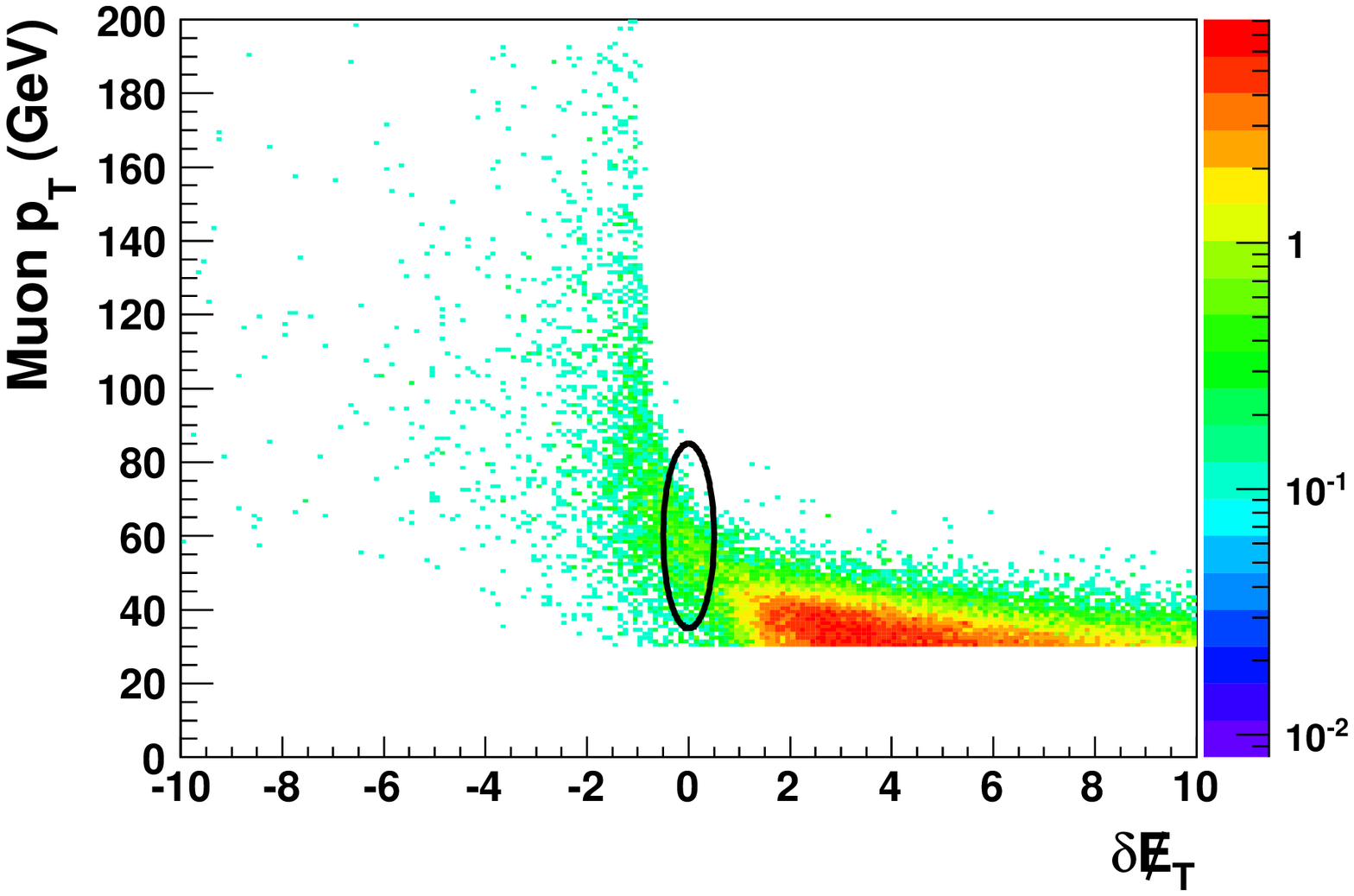} &
\includegraphics[width=8cm]{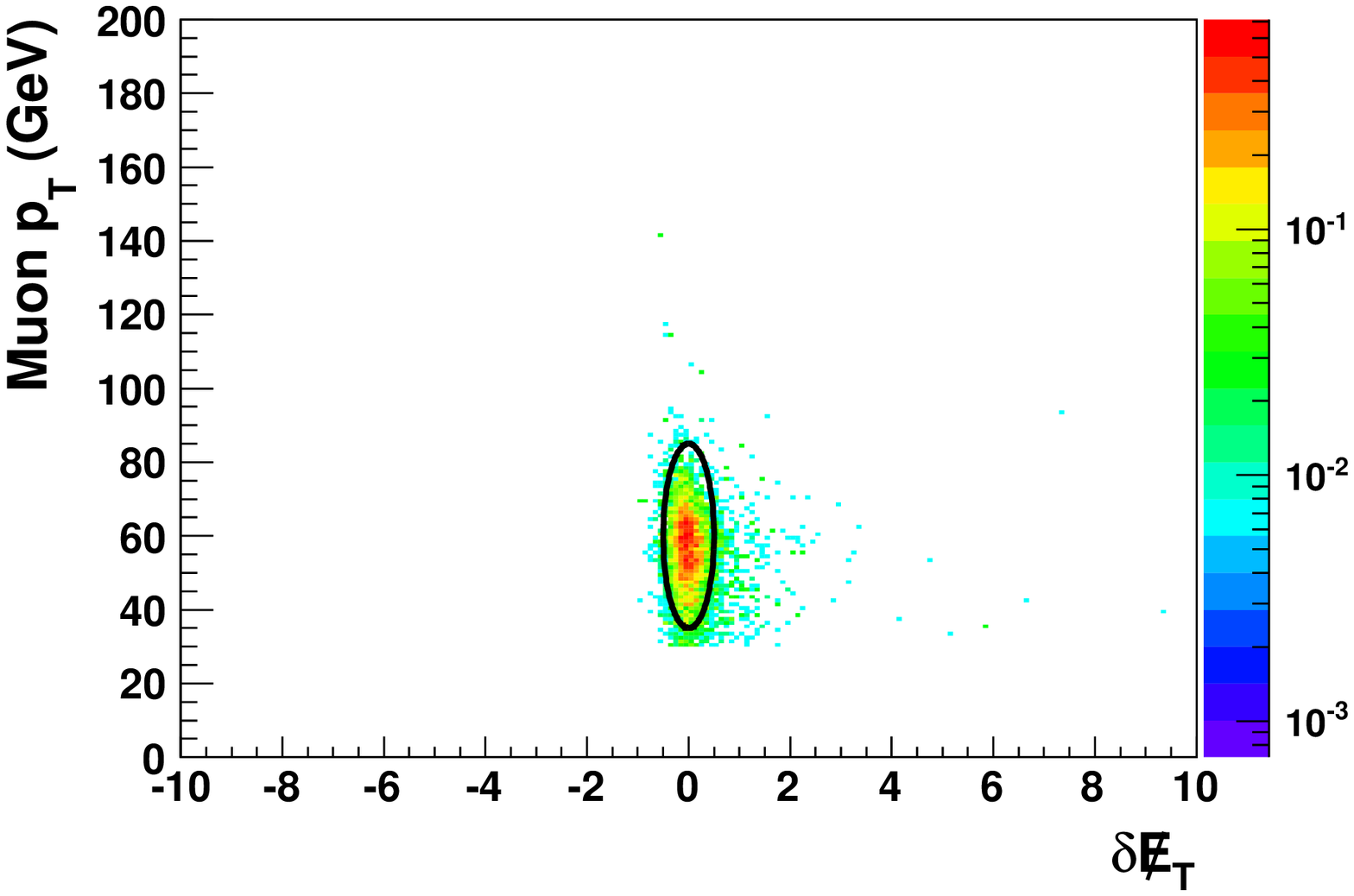} \\
\end{tabular}
\caption{
\label{fig2}
Event distribution in the $(\delta\met,p_T^\mu)$ plane for SM backgrounds (a) 
and $h \to \tau^{\pm}\mu^{\pm}$ signal (b) before applying the final selection criterium
in the $(\delta\met,p_T^\mu)$ plane, corresponding to the oval. $\delta\met$ is
defined in eqn (\ref{eqn3}). 
}
\end{center}
\end{figure}

The final selection criterium has been chosen to exploit the kinematic properties of 
$h \to \tau^{\pm} \mu^{\mp}$ compared to the remaining SM backgrounds.
The $\tau$ lepton produced in this Higgs boson decay is higly boosted. Assuming massless particles in the 
$\tau$-lepton decay $\tau \to e \bar{\nu}\nu$ and using the collinear approximation for these
decay products, the $\tau$ momentum $p_\tau$ can be expressed from the measured electron momentum $p_e$ as
\begin{eqnarray}
\label{eqn1}
p_\tau & = & \alpha p_e ~~~.
\end{eqnarray}
The momentum of the two-neutrinos system is then simply $(\alpha-1)p_e$; and its transverse component should correspond
to the missing transverse energy measured in the detector.
This $\alpha$ factor can then be related to the Higgs boson mass with the following formula:
\begin{eqnarray}
\label{eqn2}
\alpha & = & \frac{M^2_h }{4 E_e E_\mu \sin^2 \left( \theta_{e\mu}/2 \right) }
\end{eqnarray}
using the electron and muon energies $E_e$ and $E_\mu$, and the $\theta_{e\mu}$ angle between the electron
and the muon directions. For $M_h=$126~GeV, 
$\alpha$ can be used to compare the transverse momentum
of the two-neutrinos system, computed under the above hypotheses, to the measured value of transverse missing 
energy of the event ($\met^{reco}$):
\begin{eqnarray}
\label{eqn3}
\delta\met & = & \frac{(\alpha - 1) p_T^e - \met^{reco}}{\met^{reco}}
\end{eqnarray}
The distribution of the $\delta\met$ variable from Eqn.~\ref{eqn3} is shown in Fig.~\ref{fig1}a after applying all selection
criteria described so far. 
As expected, the distribution for the signal is peaked at 0
while the backgrounds have lower or higher values, 
which means that the measured value $\met^{reco}$ is not 
compatible with the value of $(\alpha - 1) p_T^e$ 
expected for a Higgs boson  of mass of 126~GeV decaying to 
$\tau^{\pm} \mu^{\mp} \to e^{\pm}\nu\bar{\nu} \mu^{\mp}$.
There are however important overlaps between signal and background 
 for this  distribution.
Therefore, we also 
exploit { that the transverse
momentum of the muon,}  $p_T^\mu$,
is higher in $h \to \tau^{\pm} \mu^{\mp}$
events  than in
SM background events (see Fig.~\ref{fig1}b). The 2-dimensional distribution $(\delta\met,p_T^\mu)$ is shown
in Fig.~\ref{fig2}. Events are then required to be in an ellipse centered around the signal distribution
and defined as:
\begin{eqnarray}
\left( \frac{p_T^\mu-60}{25} \right)^2 + \left( \frac{\delta\met}{0.25} \right) < 1
\end{eqnarray}

The number of background and signal events at each step of the selection are reported in 
Tab.~\ref{tab2} for an integrated luminosity of 20~$fb^{-1}$ at $\sqrt{s}=8$~TeV. 
A good signal to background ratio is achieved, 
 allowing  sensitivity to the 
$h\to \tau^{\pm}\mu^{\mp}$ signal with a  branching ratio below 1\%.
In this Table~\ref{tab2}, contributions from SM Higgs decays, which are included in
the total background evaluation, are solely indicated to show how the angular criteria
reduce their contribution. At the final stage, 
LFV Higgs decays largely dominates over 
SM Higgs decays, 
whose contribution is almost negligible compared to other SM backgrounds. 
The number of events selected after applying all selection criteria
are reported in Tab.~\ref{tab3}. The dominant backgrounds are $pp \to Z \to \tau^{\pm}\tau^{\mp}$
and diboson pair production (mainly $W^+W^-$). 

\begin{table}[htb]
\begin{center}
\begin{tabular}{ | l || r || r | r | r || r | r|}
\hline
Selection criteria & $N_{backgrd.}$  & $N_{h\to \tau\tau}$ & $N_{h\to WW}$ & $N_{h\to ZZ}$ & Signal efficiency (\%) & $N_{sig.}$ \\
\hline\hline
 $\geq$~1 muon with $p_T>30$~GeV and $|\eta|<2.1$ and   &  59271   $\pm$  76 &  89.   $\pm$ 3.  & 235.  $\pm$ 5.  &  4.2 $\pm$ 0.7 & 21.2  $\pm$ 0.1  & 145.4 $\pm$ 0.9  \\
 $\geq$~1 electron with $p_T>15$~GeV and $|\eta|<2.5$   & & & & & & \\
\hline
 exactly 2 OS leptons                                   &  58447   $\pm$  75 &  89.   $\pm$ 3.  & 235.  $\pm$ 5.  &  2.2 $\pm$ 0.5 & 21.2  $\pm$ 0.1  & 145.4 $\pm$ 0.9  \\   
 jet veto: no jet with $p_T>30$~GeV and $|\eta|<2.5$    &  19477   $\pm$  44 &  51.   $\pm$ 2.  & 123.  $\pm$ 3.  &  1.0 $\pm$ 0.3 & 13.1  $\pm$ 0.1  &  89.7 $\pm$ 0.7  \\
 $\Delta\phi(e,\mu) > 2.7$                              &  13261   $\pm$  36 &  40.   $\pm$ 2.  &   8.7 $\pm$ 0.9 &  0.1 $\pm$ 0.1 & 10.7  $\pm$ 0.1  &  72.9 $\pm$ 0.7  \\   
 $\Delta\phi(e,\met) < 0.3$                             &   3885   $\pm$  20 &  15.   $\pm$ 1.  &   2.4 $\pm$ 0.5 &  0.1 $\pm$ 0.1 &  7.85 $\pm$ 0.09 &  53.7 $\pm$ 0.6  \\
\hline
 2D cut in $(\delta\met,p_T^\mu)$ plane                &     53   $\pm$   2 &   0.6  $\pm$ 0.3 &   0.5 $\pm$ 0.2 &  0             &  5.34 $\pm$ 0.07 &  36.5 $\pm$ 0.5  \\ 
\hline
\end{tabular}
\caption{\label{tab2}
Selection criteria for the $h\to \tau^{\pm}\mu^{\mp}$ search at the $\sqrt{s}=8$~TeV LHC with $\mathcal{L}=20~fb^{-1}$
with the total number of events expected from SM backgrounds, the contribution of SM Higgs decay to the total background, and the signal efficiency~(\%)
and the number of signal events expected for $BR(h \to \tau^{\pm}\mu^{\mp})=10^{-2}$; uncertainties are statistical only.
}
\end{center}
\end{table}

\begin{table}[htb]
\begin{center}
\begin{tabular}{ | l | r | }
  \hline
  Processes  & Number of events \\
  \hline
  \hline
  $Z+$jets        &    34.7 $\pm$ 1.9 \\
  $t\bar{t}$      &     1.0 $\pm$ 0.3 \\
  single-top      &     0.6 $\pm$ 0.3 \\
  diboson         &    14.9 $\pm$ 1.2\\
  SM Higgs bkgrd. &     1.1 $\pm$ 0.3\\
  \hline\hline
  Total SM backgrounds     & 52.7 $\pm$ 2.3 \\
  \hline
  \hline
  $h\to \tau^{\pm}\mu^{\mp}$ signal & 36.5 $\pm$ 0.5\\
  \hline
\end{tabular}
\caption{\label{tab3}
After selection criteria of Table~\ref{tab2} for the $h\to \tau^{\pm}\mu^{\mp}$ search at the $\sqrt{s}=8$~TeV LHC with $\mathcal{L}=20~fb^{-1}$,
number of events expected for each SM background and for the signal normalized to $\sigma_{SM} (pp (gg) \to h) \times BR(h \to \tau^{\pm}\mu^{\mp})$
with $BR(h \to \tau^{\pm}\mu^{\mp})=10^{-2}$.
}
\end{center}
\end{table}

The number of events obtained in Tab.~\ref{tab3} were used to extract a limit
on $BR(h \to \tau^{\pm}\mu^{\pm})$. For that purpose, the modified frequentist
$CL_s$~\cite{cls:junkread} method was used. As in our previous study~\cite{DLV}, a 
3\% systematic uncertainty was set on the signal and background yields due to 
experimental uncertainties on lepton identifications. In addition, we also took into
account the 15\% systematic uncertainty on Higgs boson production via gluon fusion for 
the signal and for the backgrounds from SM Higgs decays.
The expected 95\% C.L. limit calculated with this procedure is:
%{\it voir commentaire apres eqn1}
\begin{eqnarray}
\label{eqn4}
\frac{\sigma(pp(gg)\to h)}{\sigma_{SM}(pp(gg)\to h)}
BR(h \to \tau^{\pm}\mu^{\pm}) < 4.5 \times 10^{-3}
\end{eqnarray}

%\newpage

%~
%\newpage

The reach indicated by eqn (\ref{eqn4}) is phenomenologically
interesting. The Cheng-Sher ans\"atz \cite{CS},
which can parametrise  
extra-dimensional models of flavour,
gives
flavour-changing couplings  of order
$y_{\tau \mu} \sim \sqrt{m_\tau m_\mu/v^2}$, where
$v = 174$ GeV is the Higgs vev. For the $h$ boson,
this gives 
\beq
BR(\Htm) \simeq \frac{m_\mu}{m_\tau} BR(\Htt)
\simeq  4.1 \times 10^{-3}
\eeq
where the last number applies when $h$ has
 the  Standard Model  branching ratio to taus. 

The sensitivity of eqn (\ref{eqn4})  can be compared to the
  prospects  of $\Htm$
  with Higgs production via VBF (where hadronic and leptonic
tau decays could be used),  as was
studied by Harnik, Kopp, and Zupan \cite{HKZ}.
  Current data indicate 
that the   production cross sections via
 gluon fusion and VBF have approximately their SM values,
implying  more signal events in the channel
$pp(gg) \to \Htm \to e^\pm \mu^\mp + \met$.   We find a
signal efficiency of 5.34 \%. 
Although the results of \cite{HKZ} are presented
differently from ours,  the gluon fusion channel
with leptonic taus, 
appears to have  better 
 sensitivity than VBF with
hadronic taus.

This analysis used  missing transverse energy 
reconstructed from the whole event. 
However, the $\met$ modeling in a fast detector
simulation is too optimistic;
a full detector 
simulation including pile-up effects
will enlarge $\met$ resolution and the overlap 
between signal and backgrounds,
thus reducing slightly the sensitivity. 
These effects have to be studied by the LHC
collaborations in order to perform this analysis.
We estimate that  the sensitivity is worsened
 by a factor $\lsim$ 2,
if the missing transverse
energy is reconstructed solely from the lepton momenta.

 The SM Higgs boson is very narrow, mostly decaying 
to $b\bar{b}$ via a coupling $y_b \simeq 1/35$. 
Its  branching ratios are
therefore a  sensitive probe of  small non-standard Higgs couplings
since, away from the  peak of the Breit-Wigner, 
observables lose the $1/y_b^2$ amplification
and are suppressed by another small Higgs coupling squared. 
This is different from the $Z$ boson, many of whose non-standard
couplings are better constrained by low energy observables.
In this paper we explored the sensitivity of  the LHC with
20~$fb^{-1}$ of data to a {SM-like} Higgs boson decaying to
$\tau^\pm \mu^\mp$ or $\tau^\pm e^\mp$. We considered
Higgs boson production via  the largest cross section,
which is gluon fusion, and leptonic tau decays,
because this gives a clean signature. Assuming 
experimental  systematic uncertainties of  3\%, and accurate
missing transverse energy  reconstruction, we find that
the LHC could exclude at 95\% C.L. $BR(\Htm),BR(\Hte) < 
4.5 \times 10^{-3}$.

\section*{Acknowledgements}

We thank  the authors of \cite{HKZ}, in particular Joachim Kopp
and Jure Zupan, for the discussions which initiated this work,
and subsequent comments and suggestions; and
we thank Sylvain Lacroix for  his contributions to  
 \cite{DLV}, upon which this analysis relies.

%%%%%%%%%%%%%%%%%%%%%%%%%%%%%%%%%%%%%%%%%%%%%%%%%%%%%%%%%%%%%%%%%%%%%%
%% SECT  %%%%%%%%%%%%%%%%%%%%%%%%%%%%%%%%%%%%%%%%%%%%%%%%%%%%%%
%%%%%%%%%%%%%%%%%%%%%%%%%%%%%%%%%%%%%%%%%%%%%%%%%%%%%%%%%%%%%%%%%%%%%%


\begin{thebibliography}{222222}




\bibitem{LHCH}
  S.~Chatrchyan {\it et al.}  [CMS Collaboration],
  ``Observation of a new boson at a mass of 125 GeV with the CMS experiment at the LHC,''
  Phys.\ Lett.\ B {\bf 716} (2012) 30
  [arXiv:1207.7235 [hep-ex]].
  %%CITATION = ARXIV:1207.7235;%%
G.~Aad {\it et al.}  [ATLAS Collaboration],
  ``Observation of a new particle in the search for the Standard Model Higgs boson with the ATLAS detector at the LHC,''
  Phys.\ Lett.\ B {\bf 716} (2012) 1
  [arXiv:1207.7214 [hep-ex]].
  %%CITATION = ARXIV:1207.7214;%%



 \bibitem{LFVHcoll}
 T.~Han and D.~Marfatia,
  ``$h \to \mu \tau$ at hadron colliders,''
  Phys.\ Rev.\ Lett.\  {\bf 86} (2001) 1442
  [hep-ph/0008141].
  %%CITATION = HEP-PH/0008141;%%
%  U.~Cotti, L.~Diaz-Cruz, C.~Pagliarone and E.~Vataga,
%  ``Search for the lepton flavor violating Higgs decay H $\to$ tau mu at hadron colliders,''
%  eConf C {\bf 010630} (2001) P102
%  [hep-ph/0111236].
  %%CITATION = HEP-PH/0111236;%%
  K.~A.~Assamagan, A.~Deandrea and P.~A.~Delsart,
  ``Search for the lepton flavor violating decay 
$A_0/H_0 \to \tau^\pm \mu^\mp$ at
  hadron colliders,''
  Phys.\ Rev.\  D {\bf 67} (2003) 035001
  [arXiv:hep-ph/0207302].
  %%CITATION = PHRVA,D67,035001;%%
S.~Arcelli,
  ``Search for H/A $\to$ mu mu and tau mu at the LHC,''
  Eur.\ Phys.\ J.\  C {\bf 33} (2004) S726.
  %%CITATION = EPHJA,C33,S726;%%







\bibitem{LFV}
  R.~Diaz, R.~Martinez and J.~A.~Rodriguez,
  ``Lepton flavor violation in the two Higgs doublet model type III,''
  Phys.\ Rev.\  D {\bf 63} (2001) 095007
  [arXiv:hep-ph/0010149].
  %%CITATION = PHRVA,D63,095007;%%
 E.~O.~Iltan,
  ``Electric dipole moments of charged leptons and lepton flavor violating interactions in the general two Higgs doublet model,''
  Phys.\ Rev.\ D {\bf 64} (2001) 013013
  [hep-ph/0101017].
  %%CITATION = HEP-PH/0101017;%%

\bibitem{anna}
A.~Brignole and A.~Rossi,
  ``Anatomy and phenomenology of mu-tau lepton flavor violation in the MSSM,''
  Nucl.\ Phys.\ B {\bf 701} (2004) 3
  [hep-ph/0404211].
  %%CITATION = HEP-PH/0404211;%%

\bibitem{ACK} 
   Abdessalem Arhrib, Yifan Cheng, Otto C. W. Kong,
``    A Comprehensive Analysis on Lepton Flavor Violating Higgs to mu + tau Decay in Supersymmetry without R Parity'',
  arXiv:1210.8241 



%\bibitem{CDF}
%  D.~Acosta {\it et al.}  [CDF Collaboration],
%  ``Search for lepton flavor violating decays of a heavy neutral particle in $p \bar{p}$ collisions at $\sqrt{s} = 1.8$ TeV,''
%  Phys.\ Rev.\ Lett.\  {\bf 91} (2003) 171602
%  [hep-ex/0307012].
%  %%CITATION = HEP-EX/0307012;%%



\bibitem{japonais}
  S.~Kanemura, T.~Ota and K.~Tsumura,
  ``Lepton flavor violation in Higgs boson decays under the rare tau decay results,''
  Phys.\ Rev.\ D {\bf 73} (2006) 016006
  [hep-ph/0505191].
  %%CITATION = HEP-PH/0505191;%%


\bibitem{gerald}
  S.~Davidson and G.~J.~Grenier,
  ``Lepton flavour violating Higgs and tau to mu gamma,''
  Phys.\ Rev.\ D {\bf 81} (2010) 095016
  [arXiv:1001.0434 [hep-ph]].
  %%CITATION = ARXIV:1001.0434;%%

\bibitem{oleg2}
  A.~Goudelis, O.~Lebedev and J.~-h.~Park,
  ``Higgs-induced lepton flavor violation,''
  Phys.\ Lett.\ B {\bf 707} (2012) 369
  [arXiv:1111.1715 [hep-ph]].
  %%CITATION = ARXIV:1111.1715;%%



\bibitem{HKZ}
  R.~Harnik, J.~Kopp and J.~Zupan,
  ``Flavor Violating Higgs Decays,''
  arXiv:1209.1397 [hep-ph].
  %%CITATION = ARXIV:1209.1397;%%

\bibitem{Chen:2006hp}
  C.~-H.~Chen and C.~-Q.~Geng,
  ``Lepton flavor violation in tau decays,''
  Phys.\ Rev.\ D {\bf 74} (2006) 035010
  [hep-ph/0605299].
  %%CITATION = HEP-PH/0605299;%%
%desintegrations vers mesons *scalaires* = bornes sur h, H


\bibitem{MPR}
  D.~McKeen, M.~Pospelov and A.~Ritz,
  ``Modified Higgs branching ratios versus CP and lepton flavor violation,''
  arXiv:1208.4597 [hep-ph].
  %%CITATION = ARXIV:1208.4597;%%

\bibitem{Ellisetal} 
  G.~Blankenburg, J.~Ellis and G.~Isidori,
  ``Flavour-Changing Decays of a 125 GeV Higgs-like Particle,''
  Phys.\ Lett.\ B {\bf 712} (2012) 386
  [arXiv:1202.5704 [hep-ph]].
  %%CITATION = ARXIV:1202.5704;%%




\bibitem{DCGM}
  J.~L.~Diaz-Cruz, D.~K.~Ghosh and S.~Moretti,
  ``Lepton Flavour Violating Heavy Higgs Decays Within the nuMSSM and Their Detection at the LHC,''
  Phys.\ Lett.\ B {\bf 679} (2009) 376
  [arXiv:0809.5158 [hep-ph]].
  %%CITATION = ARXIV:0809.5158;%%



\bibitem{BNGL}
  K.~S.~Babu and S.~Nandi,
  ``Natural fermion mass hierarchy and new signals for the Higgs boson,''
  Phys.\ Rev.\ D {\bf 62} (2000) 033002
  [hep-ph/9907213].
  %%CITATION = HEP-PH/9907213;%%
  G.~F.~Giudice and O.~Lebedev,
  ``Higgs-dependent Yukawa couplings,''
  Phys.\ Lett.\ B {\bf 665} (2008) 79
  [arXiv:0804.1753 [hep-ph]].
  %%CITATION = ARXIV:0804.1753;%%


\bibitem{hhg}
J.F.~Gunion, H.E.~Haber, G.~Kane and S.~Dawson,
{\it The Higgs Hunter's Guide} (Perseus Publishing,
Cambridge, MA, 1990).

\bibitem{ana}
  U.~Ellwanger, C.~Hugonie and A.~M.~Teixeira,
  ``The Next-to-Minimal Supersymmetric Standard Model,''
  Phys.\ Rept.\  {\bf 496} (2010) 1
  [arXiv:0910.1785 [hep-ph]].
  %%CITATION = ARXIV:0910.1785;%%

\bibitem{BEetal}
  G.~Belanger, U.~Ellwanger, J.~F.~Gunion, Y.~Jiang, S.~Kraml and J.~H.~Schwarz,
  ``Higgs Bosons at 98 and 125 GeV at LEP and the LHC,''
  arXiv:1210.1976 [hep-ph].
  %%CITATION = ARXIV:1210.1976;%%


\bibitem{DH}
  S.~Davidson and H.~E.~Haber,
  ``Basis-independent methods for the two-Higgs-doublet model,''
  Phys.\ Rev.\ D {\bf 72} (2005) 035004
   [Erratum-ibid.\ D {\bf 72} (2005) 099902]
  [hep-ph/0504050].
  %%CITATION = HEP-PH/0504050;%%

\bibitem{paradisi}
  P.~Paradisi,
  ``Higgs-mediated $\tau\to \mu$  and $\tau \to e$ transitions in II Higgs doublet model and supersymmetry,''
  JHEP {\bf 0602} (2006) 050
  [hep-ph/0508054].
  %%CITATION = HEP-PH/0508054;%%



\bibitem{HxsecWG}
  LHC Higgs Cross Section Working Group, A.~David, A.~Denner, M.~Duehrssen, M.~Grazzini, C.~Grojean, G.~Passarino and M.~Schumacher {\it et al.},
  ``LHC HXSWG interim recommendations to explore the coupling structure of a Higgs-like particle,''
  arXiv:1209.0040 [hep-ph].
  %%CITATION = ARXIV:1209.0040;%%


\bibitem{aldo}
  G.~Cacciapaglia, A.~Deandrea, G.~D.~La Rochelle and J.~-B.~Flament,
  ``Higgs couplings beyond the Standard Model,''
  arXiv:1210.8120 [hep-ph].
  %%CITATION = ARXIV:1210.8120;%%


\bibitem{Sjostrand:2007gs}
  T.~Sjostrand, S.~Mrenna and P.~Z.~Skands,
  ``A Brief Introduction to PYTHIA 8.1,''
  Comput.\ Phys.\ Commun.\  {\bf 178} (2008) 852
  [arXiv:0710.3820 [hep-ph]].
  %%CITATION = ARXIV:0710.3820;%%

\bibitem{DLV}
  S.~Davidson, S.~Lacroix and P.~Verdier,
  ``LHC sensitivity to lepton flavour violating Z boson decays,''
  JHEP {\bf 1209} (2012) 092
  [arXiv:1207.4894 [hep-ph]].
  %%CITATION = ARXIV:1207.4894;%%



\bibitem{Dittmaier:2011ti} 
  S.~Dittmaier {\it et al.}  [LHC Higgs Cross Section Working Group Collaboration],
  ``Handbook of LHC Higgs Cross Sections: 1. Inclusive Observables,''
  arXiv:1101.0593 [hep-ph].
  %%CITATION = ARXIV:1101.0593;%%

\bibitem{Ovyn:2009tx}
  S.~Ovyn, X.~Rouby and V.~Lemaitre,
  ``DELPHES, a framework for fast simulation of a generic collider experiment,''
  arXiv:0903.2225 [hep-ph].
  %%CITATION = ARXIV:0903.2225;%%

\bibitem{Cacciari:2011ma}
  M.~Cacciari, G.~P.~Salam and G.~Soyez,
  ``FastJet user manual,''
  Eur.\ Phys.\ J.\ C {\bf 72} (2012) 1896
  [arXiv:1111.6097 [hep-ph]].
  %%CITATION = ARXIV:1111.6097;%%


\bibitem{cls:junkread}
  T.~Junk,
  ``Confidence level computation for combining searches with small statistics,''
  Nucl.\ Instrum.\ Meth.\ A {\bf 434} (1999) 435
  [hep-ex/9902006];
  %%CITATION = HEP-EX/9902006;%%

  A. Read,
  ``Modified Frequentist Analysis of Search Results (The CLs Method),''
  CERN-2000-005 (2000).



\bibitem{CS}
  T.~P.~Cheng and M.~Sher,
  ``Mass Matrix Ansatz and Flavor Nonconservation in Models with Multiple Higgs Doublets,''
  Phys.\ Rev.\ D {\bf 35} (1987) 3484.
  %%CITATION = PHRVA,D35,3484;%%







\end{thebibliography}
\end{document}